# Intrinsic p-type W-based transition metal dichalcogenide by substitutional Ta-doping

Yajun Fu[1,a)], Mingsheng Long[1,a)], Anyuan Gao[1], Yu Wang[1], Chen Pan[1], Xiaowei Liu[1], Junwen Zeng[1], Kang Xu[1], Lili Zhang[1], Erfu Liu[1], Weida Hu[2], Xiaomu Wang[3] & Feng Miao[1,b)]

[1]National Laboratory of Solid State Microstructures, School of Physics, Collaborative Innovation Center of Advanced Microstructures, Nanjing University, Nanjing 210093, China.

[2]National Laboratory for Infrared Physics, Shanghai Institute of Technical Physics, Chinese Academy of Sciences, Shanghai 200083, China.

[3]School of Electronic Science and Technology, Nanjing University, Nanjing 210093, China.

[a)] Y. Fu and M. Long contributed equally to this work.

[b)]Author to whom correspondence should be addressed. Electronic mail: miao@nju.edu.cn.

**Two-dimensional (2D) transition metal dichalcogenides (TMDs) have recently emerged as promising candidates for future electronics and optoelectronics. While most of TMDs are intrinsic n-type semiconductors due to electron donating which originates from chalcogen vacancies, obtaining intrinsic high-quality p-type semiconducting TMDs has been challenging. Here, we report an experimental approach to obtain intrinsic p-type Tungsten (W)-based TMDs by substitutional Ta-doping. The obtained few-layer Ta-doped WSe$_2$ (Ta$_{0.01}$W$_{0.99}$Se$_2$) field-effect transistor (FET) devices exhibit competitive p-type**

performances, including ~$10^6$ current on/off at room temperature. We also demonstrate high quality van der Waals (vdW) p-n heterojunctions based on $Ta_{0.01}W_{0.99}Se_2/MoS_2$ structure, which exhibit nearly ideal diode characteristics (with an ideality factor approaching 1 and a rectification ratio up to $1\times10^5$) and excellent photodetecting performance. Our study suggests that substitutional Ta-doping holds great promise to realize intrinsic p-type W-based TMDs for future electronic and photonic applications.

Since the recent emergence of transition metal dichalcogenides (TMDs), tremendous efforts have been involved to explore their interesting physical phenomena and potential applications.[1-6] Due to their moderate band gap and strong light-matter coupling,[7, 8] two-dimensional (2D) TMDs are top material candidates for the next generation electronic and photonic devices.[9-11] Like traditional complementary metal-oxide-semiconductor (CMOS) electronics, both p- and n-type TMDs are indispensable for realizing complementary device applications. However, most of semiconducting TMDs are intrinsic n-type originating from chalcogen vacancies. Although engineering extrinsic factors could help achieving p-type behaviors, such as metal work-function engineering,[12] ionic liquid gating,[13, 14] and chemical doping[15], obtaining intrinsic high quality p-type semiconducting TMDs has been challenging.

On the other hand, as a representative TMD material, tungsten diselenide ($WSe_2$) could be a good candidate to realize an intrinsic p-type semiconductor due to the observed ambipolor behavior[16, 17] and its relatively high hole mobility which benefits the device performances.[12, 18, 19] Moreover, due to its unique electronic band structure and strong spin-orbit interaction (originating from W as a heavy 5d element), the spin degeneracy near the valence band edge of $WSe_2$ is split by $\sim 400$ meV, much stronger than that of $MoS_2$ ($\sim 160$ meV).[20, 21] This results in many intriguing physical phenomena, such as valley pseudospins,[22] electric-field-driven Zeeman-type spin splitting,[23] and negative electronic compressibility.[24, 25] Therefore, considering the small effective hole mass ($0.45m_0$),[24] high hole mobility, as well as the unique band structure, an intrinsic high-quality p-type $WSe_2$ could provide an excellent platform for exploring novel physics and device applications.

In this work, we report the realization of p-type $WSe_2$ ($Ta_{0.01}W_{0.99}Se_2$) by substitutional Ta-doping. We perform various material characterizations with results indicating successful doping of Ta element. The few-layer $Ta_{0.01}W_{0.99}Se_2$ FETs show competitive p-type performances including high current on/off ratio up to $10^6$. We further fabricate vdW p-n heterojunctions by vertically stacking $Ta_{0.01}W_{0.99}Se_2$ and few-layer $MoS_2$. Nearly ideal diode behaviors (with an ideality factor approaching 1 and a rectification ratio of $1 \times 10^5$) and high efficiency photoresponce were observed, suggesting high quality of our intrinsic p-type $WSe_2$ samples. Our results pave the way for future electronic and optoelectronic applications based on complementary TMD materials.

Elemental doping is an efficient and widely adopted method to adjust both carrier type and carrier density in traditional semiconductors. Recently, elemental doping method has been utilized to modulate the carrier type of Molybdenum (Mo)-based TMDs, resulting in intrinsic p-type characteristics.[26, 27] A feasible scenario to achieve p-type $WSe_2$ by element doping is to replace certain W atoms (six valence electrons) by Tantalum (Ta) atoms with five valence electrons and matched atomic radius.

Figure 1a schematically shows the crystalline structure of Ta-doped $WSe_2$, where Ta atoms replace W atoms randomly. The Ta-doped $WSe_2$ (nominal value of Ta in atomic ratio is ~1%) bulk single crystals were grown by chemical vapor transport (CVT) method [28-30] (see details in experimental section). We identified the crystal phase by utilizing X-ray diffraction (XRD) in a powder mode (Figure 1b). The peak positions for Ta-doped and pristine $WSe_2$ are almost identical and consistent with the standard PDF card #38-1388, indicating that good crystallinity (2H phase) was maintained in the Ta-doped $WSe_2$. According to the XRD diffraction angle data, the lattice constant along $c$ axis was calculated to be 12.977 Å and 12.960 Å for pristine and Ta-doped $WSe_2$ respectively. The in-plane constant $a$ maintains almost the same with results calculated to be 3.289 Å and 3.291 Å for pristine and Ta-doped $WSe_2$ respectively. Such results suggest that certain W atoms are successfully replaced by Ta atoms, in good consistency with prior work for Nb-doped $MoSe_2$.[27] The Raman spectra of two different types of $WSe_2$ (monolayer flakes) are shown in Figure 1c, exhibiting similar characteristics. The main peak around 250 cm$^{-1}$ is associated with both $A_{1g}$ and $E_{2g}^1$ modes (with small frequency difference). The small features located around 260, 360, 373 and 395 cm$^{-1}$ correspond to the second-order Raman processes.[31, 32] We noticed that the main peak ($A_{1g}$ and $E_{2g}^1$ modes) of Ta-doped $WSe_2$ is slightly blueshifted with ~1.4 cm$^{-1}$, which could be induced by hole doping induced phonon hardening.[33]

We further characterized the Ta element doping by utilizing X-ray photoelectron spectroscopy (XPS) and Energy Dispersive X-ray spectroscopy (EDX). XPS spectra clearly show the existence of Ta element in the doped $WSe_2$ in spite of weak feature of the Ta 4f peak (see Figure S1 in the supplementary material). According to the EDX data shown in Figure 1d, the Fermi level of Ta-doped $WSe_2$ is downshifted by ~0.4 eV while the shape feature of the valence band keeps consistent, indicating that the Fermi level of the doped samples is much closer to the valence band maximum. At the same time, W 4f and Se 3d peaks in XPS spectra of Ta-doped $WSe_2$ are upshifted by ~0.4 eV as well (see Figure S1 in the supplementary material). Although the peak feature of Ta 4f in XPS data is too weak for the calculation of the atomic ratio for each element, a value of ~0.7 at% Ta was obtained for $Ta_{0.01}W_{0.99}Se_2$ samples according to the EDX data (see Figure S2 in the supplementary material), close to the nominal elemental composition. Overall, these results clearly demonstrate that certain W atoms in $WSe_2$ single crystals are successfully substituted by Ta atoms, resulting in intrinsic hole doping.

We now focus on the electronic properties of the few-layer $Ta_{0.01}W_{0.99}Se_2$. Thin flakes (thickness below 5 nm) were exfoliated onto degenerately doped Si wafers covered with 300 nm $SiO_2$ for device fabrications. For a typical few-layer $Ta_{0.01}W_{0.99}Se_2$ device (3.4 nm), the source-drain current $I_{ds}$ exhibits good linear dependence on the bias $V_{ds}$ (within the range of ±2 V) under various gate voltages $V_{bg}$ (inset of Figure 2a), which implies the ohmic contact behavior for the hole transport at room temperature (300 K). The temperature dependent four-terminal resistance measurements were carried out, where the resistance increases monotonically with decreasing temperature, indicating a semiconducting transport behavior (as shown in Figure 2a). Additionally, the FET transfer curves measured from 25 to 300 K show excellent p-type FET characteristics when the back gate swept between -100 and +100 V (as shown in Figure 2b). The current on/off ratio exceeded $10^6$ at room temperature, which fulfills the requirement of logic applications. We noticed that a weak n-type current for high positive gate biases appeared at lower temperatures. At room temperature, the subthreshold swing is about 1 V dec$^{-1}$ for 300 nm back gate dielectric (see Figure S3 in the supplementary material), which could be further improved by utilizing thinner high-κ dielectric materials. The field-effect mobility were extracted from the linear regime of the transfer curves (see Figure S3 in the supplementary material), which reached ~16.5 cm$^2$ V$^{-1}$ s$^{-1}$ at room temperature. According to the temperature dependent transfer characteristics (Figure 2b), we found the data fit well to a hopping transport model (see Figure S4 in the supplementary material).[34] Similar to previous studies in MoS$_2$, the charge transport in Ta$_{0.01}$W$_{0.99}$Se$_2$ is thermally activated and switched from the nearest hopping to the variable range hopping at lower temperature regime.[34, 35] We also performed control experiments to study the FET properties of pristine WSe$_2$ (without Ta dopants) few layer flakes, and indeed observed an ambipolar behavior (see Figure S5 in the supplementary material).

Successful demonstration of intrinsic p-type WSe$_2$ makes it feasible to realize high-quality p-n heterojunctions. Figure 2c shows the typical gate tunable rectifying behavior at room temperature for a vertical stacked Ta$_{0.01}$W$_{0.99}$Se$_2$ (2.3 nm) and n-MoS$_2$ (2.7 nm) vdW heterojunction (with Ti/Au electrodes), in which the forward current at positive source-drain voltage increases with increasing $V_{bg}$. The rectification ratio, which is defined as the forward/reverse current ratio at the same source-drain bias, exceeded 10$^3$ for all measured gate voltages ($V_{ds} = \pm 2$ V, as shown in the inset of Figure 2c). The extracted rectification ratios ($V_{ds} = \pm 2$ V) at different gate biases are shown in Figure 2d, where the highest rectification ratio achieved is up to 1×10$^5$ at $V_{bg} = 0$ V. In addition, we fitted the rectification curves of the heterojunction device to the Shockley diode equation[36] and obtained an ideality factor ($n$) of ~1 (inset of Figure 2d), which suggests an ideal diode behavior.

Due to the excellent diode behavior, the $Ta_{0.01}W_{0.99}Se_2/MoS_2$ p-n heterojunction could result in a pronounced photovoltaic effect. The photoresponse data of our p-n junction exposed to a focused laser beam with various excitation powers (at a fixed laser wavelength $\lambda = 637$ nm) are shown in Figure 3a. And the extracted short circuit current ($I_{sc}$) and open circuit voltage ($V_{oc}$) are shown in Figure 3b. The highest $V_{oc}$ obtained here is ~ 0.5 V. The extracted photoresponsivity ($R$) and external quantum efficiency (EQE) increase while decreasing the laser power and reach values of up to ~95 mA $W^{-1}$ and ~19% ($P_{Laser} = 13.6$ μW), respectively (see Figure S6 in the supplementary material). Moreover, similar photovoltaic behavior was also observed in vdW homojunction built by vertical stacking of Ta-doped $WSe_2$ and pristine $WSe_2$ (see Figure S7 in the supplementary material).

The speed of photoresponse for photodetection devices is an important parameter to assess the device performance. A rise time of 38.9 μs and a fall time of 23.5 μs were observed (as displayed in Figure 3c), which are almost three orders of magnitude faster than the double gated $WSe_2$ lateral p-n junction (~10 ms),[37] and comparable to the p-g-n heterostructure.[38] For TMDs van der Waals heterostructures, the ultrafast charge transfer and interlayer inelastic tunneling process promote the separation of electron-hole pairs at the interface.[4, 39] Therefore, taking into account the strong built-in electric field in vertical vdW heterojunctions, the high speed of photoresponse is attribute to the efficient charge separation at the vertical heterostructure interface.

We further carried out the photocurrent mapping measurements on the devices. Figure 4a shows the measured heterojuntion, where the TMDs thin flakes are highlighted by dotted line with different colors, i.e. white and red colors represent $Ta_{0.01}W_{0.99}Se_2$ (2.9 nm) and $MoS_2$ (2.7 nm) respectively. The photocurrent mapping measurements were performed under ambient conditions, by using a 637 nm laser with power of ~269 µW. Figure 4b shows the photocurrent mapping at $V_{ds} = 0$V and $V_{bg} = 0$ V, where the most pronounced photoresponse appears within the overlapped region of $Ta_{0.01}W_{0.99}Se_2$ and $MoS_2$, originating from the formed atomically thin p-n junctions.

In summary, we have demonstrated the realization of intrinsic p-type W-based TMD by substitutional Ta-doping. The fabricated $Ta_{0.01}W_{0.99}Se_2$ FET devices exhibit competitive p-type performance, with current on/off ratio reached $10^6$. By vertical stacking $Ta_{0.01}W_{0.99}Se_2$ and few-layer $MoS_2$, the formed vdW p-n heterojunctions show excellent diode and photoresponse characteristics. Our results enrich the understanding of substitutional doped p-type TMDs, and also provide a new platform to explore novel physics and device applications.

**Methods**

*Material Synthesis*: Ta-doped and pristine $WSe_2$ bulk single crystals were synthesized by using the chemical vapor transport (CVT) method with iodine as a transporting agent. Ta, W, and Se powders were stoichiometrically mixed with an atomic ratio of Ta : W : Se = x : 1-x : 2 (x = 0, 0.005, 0.01, 0.02). The powder mixtures were pressed in tablets and sealed in quartz tubes under vacuum ($10^{-5}$ torr), followed by two weeks heat treatment in a two zone furnace where the source and growth zones of the quartz tube were kept at 980°C and 850 °C respectively. Then the furnace was cooled down to room temperature within 5 hours. The yielding single-crystals could be found in cold end of the quartz tube.

*Characterization:* The crystal structure of Ta-doped $WSe_2$ flakes were determined by X-ray diffraction (X' TRA) with Cu Kα radiation ($\lambda$ = 1.54059 Å). X-ray photoelectron spectroscopy (XPS, PHI 5000 VersaProbe) measurement was performed in ultrahigh vacuum chamber ($\leq 6.7 \times 10^{-10}$ mbar), with an Al Kα monochromatic photon source. The Energy Dispersive X-ray spectroscopy (EDX) measurement was carried out in a scanning electron microscope (SEM, S-3400N II) equipped with an X-ray Energy Dispersive Detector (EX-250). The thickness of $WSe_2$ and $MoS_2$ thin flakes were determined by using a Bruker Multimode 8 atomic force microscope (AFM). Raman spectra measurements were carried out by utilizing the Raman spectrometer with a 514.5 nm wavelength laser (under ambient conditions).

*Device Fabrication and Measurement*: The $WSe_2$ and $MoS_2$ thin flakes on degenerately doped silicon wafers (covered by 300 nm $SiO_2$) were fabricated by standard mechanical exfoliation method. The vertical heterojunctions (and homojunctions) were fabricated by using the method developed by C. R. Dean *et al*.[40] To fabricate electrical devices, electron-beam lithography process (FEI F50 with an NPGS pattern generation system) was used to design the electrodes patterns, and electron-beam evaporation was utilized to evaporate electrodes metals (typically 5 nm

Ti/40 nm Au). The variable-temperature measurements were carried out in an Oxford Instruments TeslatronTM$^{CF}$ cryostat. All the current signals were detected by Keithley 2636A dual channel digital source meter.

**Supplementary material**

Supplementary material can be found in Supporting Data in our Author Resource Center.

**Acknowledgments**

M. L., Y. F. and F. M. conceived the project and designed the experiments. Y. F., M. L., Y. W., A. G., C. P. and P. W. performed device fabrication and characterization. M. L., Y. F., X. W. and F. M. performed data analysis and interpretation. Y. F., M. L., F. M. and X. W. co-wrote the paper, and all authors contributed to the discussion and preparation of the manuscript. Y. F. and M. L. contributed equally to this work. This work was supported in part by the National Key Basic Research Program of China (2015CB921600, 2013CBA01603, 2013CB632700), the National Natural Science Foundation of China (11374142, 11322441, 61574076), the Natural Science Foundation of Jiangsu Province (BK20140017, BK20150055), the Fund of the Shanghai Science and Technology Foundation (14JC1406400), the Specialized Research Fund for the Doctoral Program of Higher Education (20130091120040), and Fundamental Research Funds for the Central Universities and the Collaborative Innovation Center of Advanced Microstructures.

**Competing financial interests**

The authors declare that they have no competing financial interests.

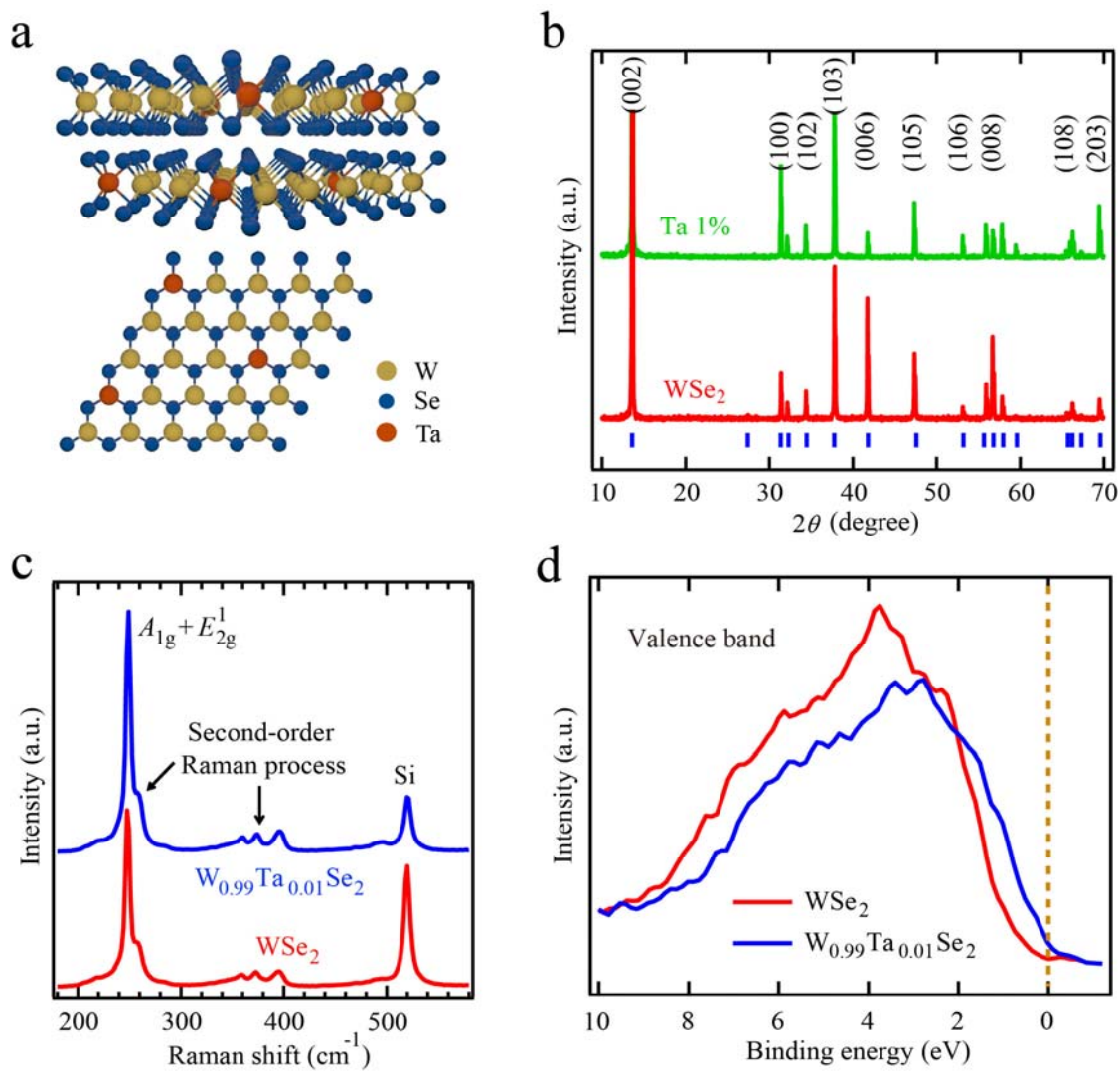

**Figure 1.** Characterization of Ta-doped $WSe_2$. a) Schematic diagrams of the Ta-doped $WSe_2$. The dopant Ta atoms randomly replace W atoms results in a hole doped $WSe_2$. b) XRD patterns for Ta-doped and pristine $WSe_2$. The little blue sticks at the bottom present the peak position of $WSe_2$ crystal from the standard PDF card #38-1388. c) Raman spectra of exfoliated monolayer Ta-doped and pristine $WSe_2$ flakes. d) Valance band spectra of Ta-doped and pristine $WSe_2$ measured by XPS. The dashed line marks the location of fermi level.

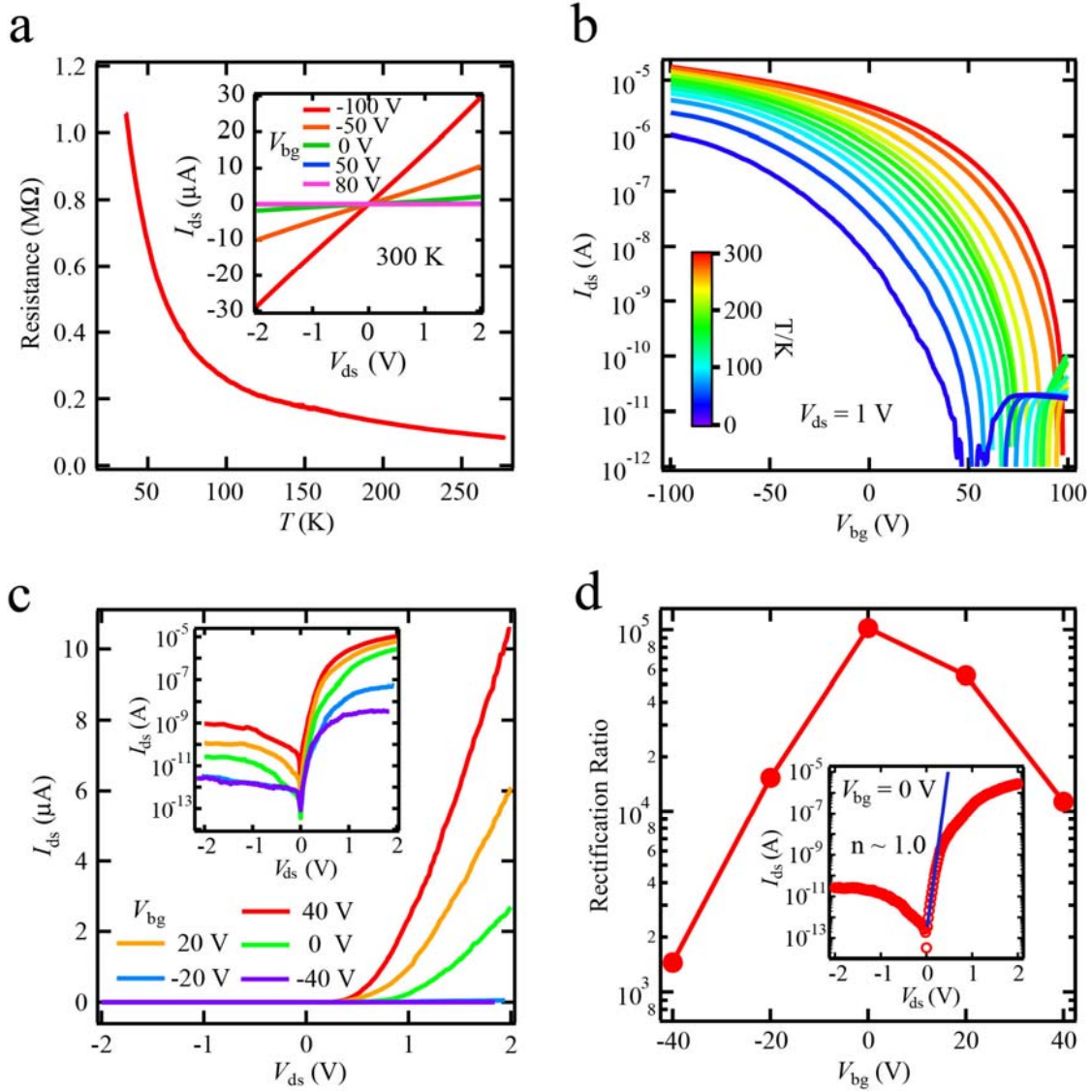

**Figure 2.** FETs and p-n heterojunctions based on $Ta_{0.01}W_{0.99}Se_2$. a) Temperature dependent resistance of $Ta_{0.01}W_{0.99}Se_2$ measured by four terminal geometry. Inset shows the $I_{ds}$-$V_{ds}$ curves for different back gate bias at 300K, the linear character suggests good contact. b) The temperature dependent Logarithmic-scale transfer curves of a $Ta_{0.01}W_{0.99}Se_2$ FET. The temperature changed from 25 K to 300 K with a step of 25 K. c) Gate tunable rectifying behavior of a typical p-n heterojunction build by vertical stacking The temperature dependent Logarithmic-scale transfer curves of a $Ta_{0.01}W_{0.99}Se_2$ and $MoS_2$. Inset: Semilogarithmic plot of the $I_{ds}$-$V_{ds}$ curves. d) The rectification ratio (extracted from the inset of (c)) as a function of gate bias. Inset: Semilogarithmic plot of the $I$-$V$ curve for $V_{bg} = 0$ V. The blue line shows linear fit to the diode equation and obtained an ideality factor of $n \sim 1$.

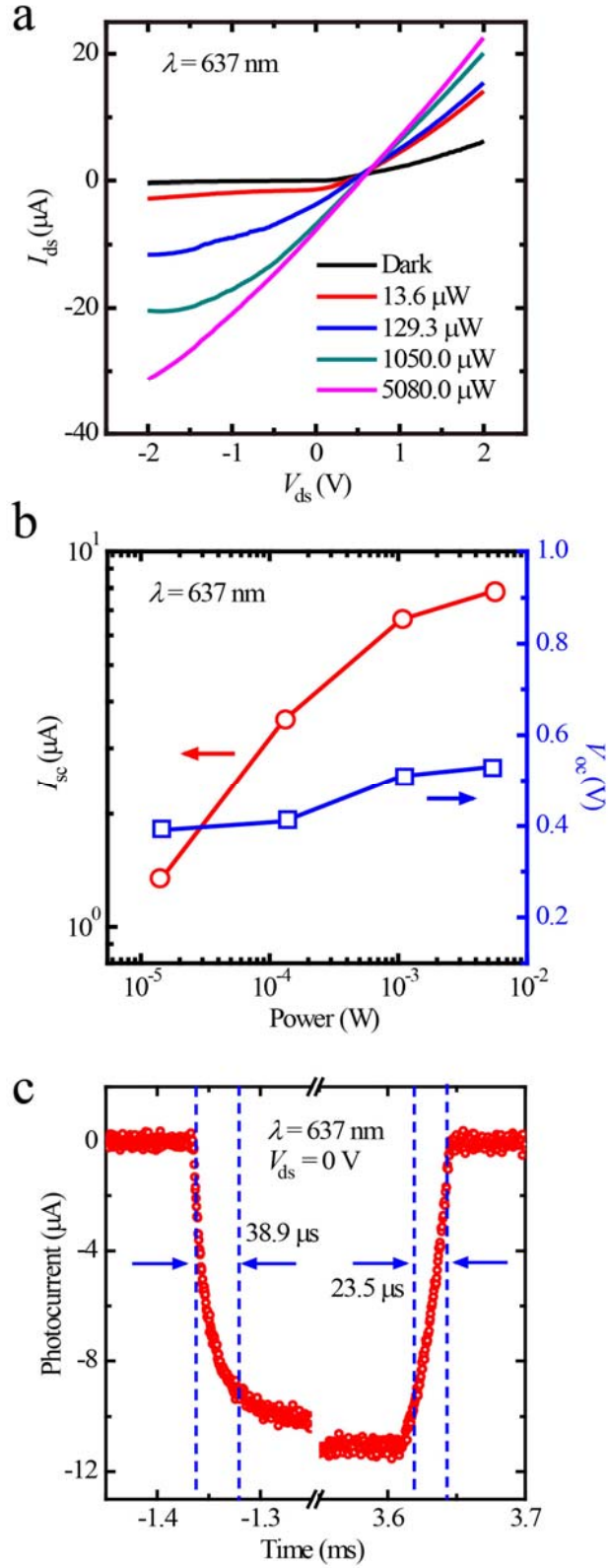

**Figure 3.** Photoresponse of a typical Ta$_{0.01}$W$_{0.99}$Se$_2$/MoS$_2$ p-n heterojunction. a) $I_{ds}$-$V_{ds}$ curves measured by exposing the p-n diode to a focused laser beam (637 nm)

with different power intensity. b) Short-circuit current ($I_{sc}$) and open-circuit voltage ($V_{oc}$) extracted from (a) as a function of various incident power intensity. c) Fast photoresponse of the p-n junction measured by using a laser (637 nm) of 1000 Hz frequency. Here the rise/fall time was defined as the photocurrent increased/decreased from 10/90% to 90/10% of the stable photocurrent.

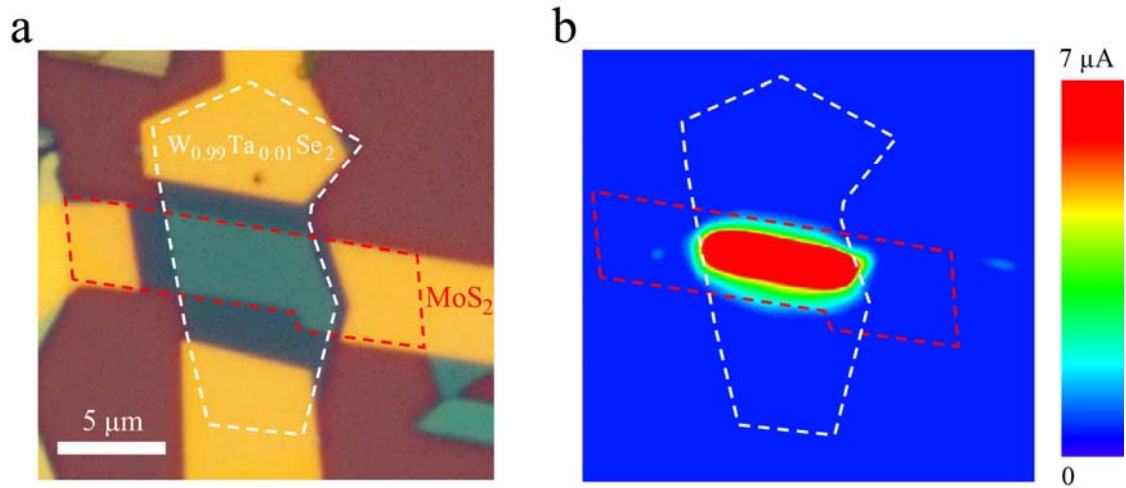

**Figure 4.** Photocurrent mapping of a typical $Ta_{0.01}W_{0.99}Se_2/MoS_2$ p-n heterojunction. a) Optical micrograph of the measured device, the scale bar is 5 μm. $MoS_2$ and Ta-doped $WSe_2$ are highlighted by red and white dashed lines, respectively. b) Photocurrent mapping of the device shown in (a) at $V_{ds}$ = 0 V and $V_{bg}$ = 0 V. The measurements were performed under ambient conditions with a laser (637 nm) power intensity of 269 μW. The same dashed line design as in (a) are guides to the eye. The photocurrent mapping unambiguously shows the photoresponse originates from the overlapped region (junction region) of $MoS_2$ and $Ta_{0.01}W_{0.99}Se_2$.

**Supplementary Material supporting the main manuscript**

# Intrinsic p-type W-based transition metal dichalcogenide by substitutional Ta-doping

Yajun Fu[1,a)], Mingsheng Long[1,a)], Anyuan Gao[1], Yu Wang[1], Chen Pan[1], Xiaowei Liu[1], Junwen Zeng[1], Kang Xu[1], Lili Zhang[1], Erfu Liu[1], Weida Hu[2], Xiaomu Wang[3] & Feng Miao[1,b)]

[1]National Laboratory of Solid State Microstructures, School of Physics, Collaborative Innovation Center of Advanced Microstructures, Nanjing University, Nanjing 210093, China.

[2]National Laboratory for Infrared Physics, Shanghai Institute of Technical Physics, Chinese Academy of Sciences, Shanghai 200083, China.

[3]School of Electronic Science and Technology, Nanjing University, Nanjing 210093, China.

Electronic mail: miao@nju.edu.cn.

**I.  XPS characterization of Ta-doped $WSe_2$.**

**II.  EDX results of $Ta_{0.01}W_{0.99}Se_2$.**

**III.  Transport properties of $Ta_{0.01}W_{0.99}Se_2$ few layer device.**

**VI.  FET property of pristine $WSe_2$ few layer device.**

**V.  Photoresponsivity and external quantum efficiency of the $Ta_{0.01}W_{0.99}Se_2/MoS_2$ p-n heterojunction.**

**VI.  Photoresponse of homojunction based on $Ta_{0.01}W_{0.99}Se_2$.**

## I. XPS characterization of Ta-doped WSe₂.

To investigate the elemental doping effect, we performed X-ray photoelectron spectroscopy (XPS) Ta-doped WSe₂ and pristine WSe₂. The survey spectra of the two types of WSe₂ show nearly identical peak features (**Figure S1**a). We further measured the spectrum of each element, as shown in Figure S1b, c, and d. The appearance of Ta 4f peak in $Ta_{0.01}W_{0.99}Se_2$ samples (Figure S1b) indicates that Ta element is successfully introduced into the sample, while no Ta-related peak has been observed in pristineWSe₂. As a result, the peaks of W 4f and Se 3d in Ta-doped WSe₂ are upshifted by ~ 0.4 eV compared to those in pristine WSe₂, as well as the valence band (shown in main text Figure 1d). However, due to the minute quantity of doped atoms, we cannot get a good peak profile of Ta 4f to allow us calculate the content of Ta atoms.

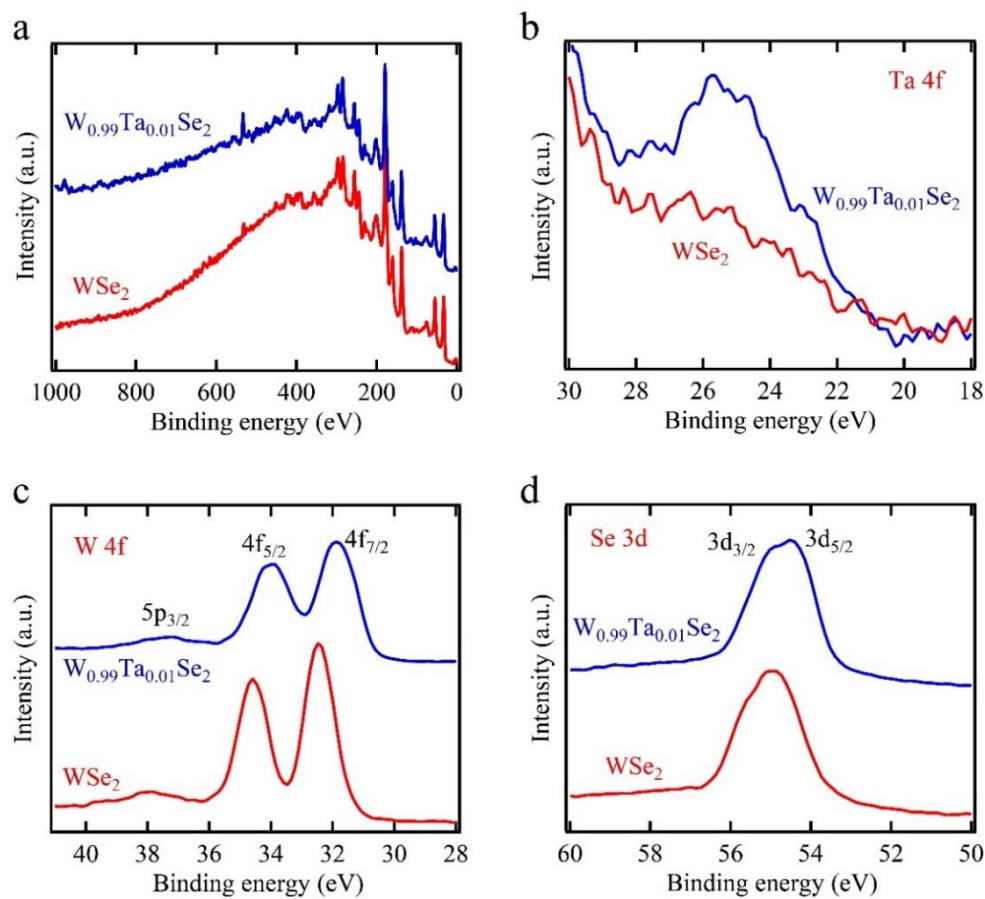

**Figure S1.** XPS characterization of Ta-doped WSe₂ and pristine WSe₂. a) XPS survey scan the binding energy in the range of 0 to 1000 eV. b, c, d) show the XPS spectra for Ta 4f, W 4f, and Se 3d, respectively.

## II. EDX results of $Ta_{0.01}W_{0.99}Se_2$.

To further determine the elemental composition of Ta-doped $WSe_2$, Energy Dispersive X-ray spectroscopy (EDX) measurements were performed. We exfoliated the bulk Ta-doped $WSe_2$ crystal onto silicon wafers (covered with 300 nm $SiO_2$), and chose thick crystal flakes (tens of nanometer, as shown in **Figure S2**a) for the measurement. A value of 0.7 at% Ta was obtained for $Ta_{0.01}W_{0.99}Se_2$ sample, as shown in Figure S2b.

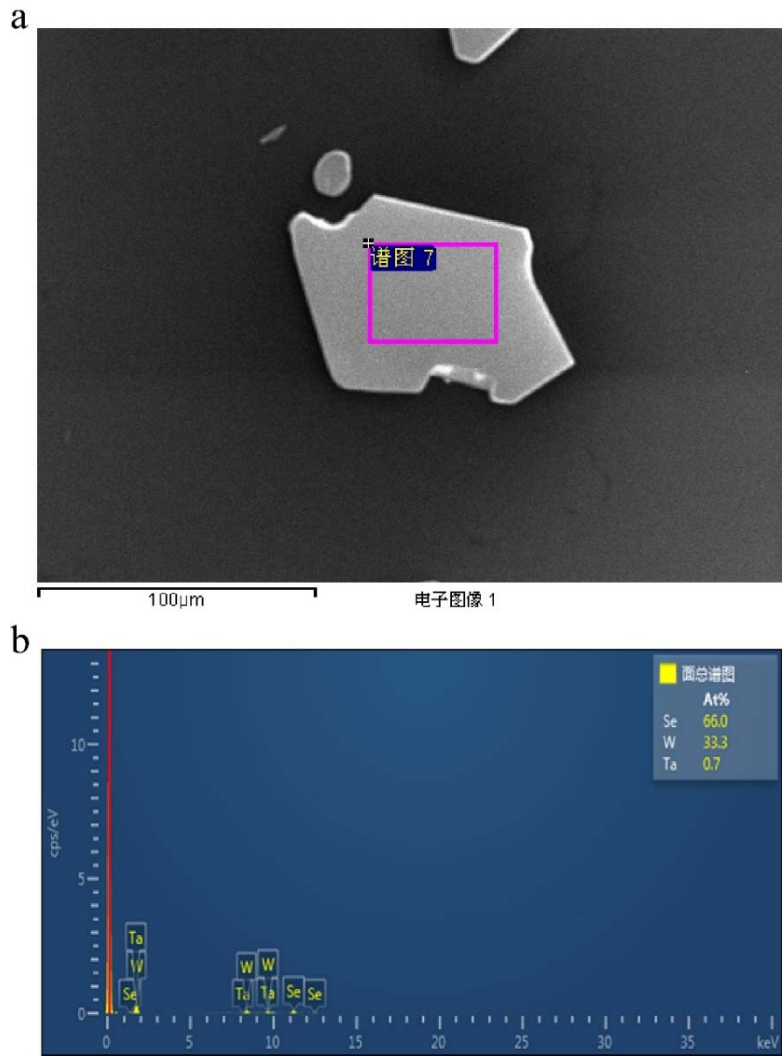

**Figure S2.** Energy Dispersive X-ray spectroscopy (EDX) measurement of Ta-doped $WSe_2$. a) The scanning electron microscope (SEM) image of a $Ta_{0.01}W_{0.99}Se_2$ crystal flake, the pink rectangle indicates the area for EDX measurement. a) EDX result for the sample shown in (a).

## III. Transport properties of $Ta_{0.01}W_{0.99}Se_2$ few layer device.

We measured the transfer characteristics of a few layer device (3.4 nm) at different temperatures (25 to 300 K), as shown in **figure S3**a. The clear p-type FET behaviors were observed as back gate bias changed from -100 to +100 V. At room temperature, an on/off ratio exceeding $10^6$ and a subthreshold swing (S) approaching 1 V dec$^{-1}$ (for 300 nm $SiO_2$ dielectric) were observed, as shown in figure S3b. Figure S3c shows $I_{ds}$-$V_{ds}$ curves measured at different temperatures (25 to 300 K) with $V_{bg}$ = -50 V. The linear behaviors for temperatures above 100 K indicate the good contact at higher temperatures. We extracted the field-effect mobility ($\mu$) at different temperatures by fitting the linear part of the transfer curves in figure S3a using the expression:

$$\mu = [dI_{ds}/dV_{bg}] \times [L/(WC_iV_{ds})]$$

where $L$ is the channel length, $W$ is the channel width, and $C_i$ is the capacitance per unit area (with 300 nm $SiO_2$ dielectric). For temperature below 275 K, the mobility decreases with lowering temperature (Figure S3d), which is consistent with transport dominated by scattering from charged impurities.[1] While for 300 K, the mobility is lower than the highest value (16.5 cm$^2$ V$^{-1}$ s$^{-1}$) at 275 K. The mobility decrease behavior at higher temperatures suggests that the electron-phonon scattering dominated the transport.[1, 2]

To further study the transport behavior, we present the conductance ($G=I_{ds}/V_{ds}$) at different $V_{bg}$ as Arrhenius plot in **Figure S4**a. The conductance in high-temperature regime (100 K to 300 K) show thermally activated transport behaviors, which can be depicted by:

$$G = G_0e^{-E_a/k_BT}$$

where $E_a$ is the activation energy, $k_B$ is the Boltzmann constant and $G_0$ is a temperature-dependent parameter extracted from the fitting curves.[3] The good fitting to the equation suggests the charge transport is thermally activated and dominated by nearest-neighbour hopping at high temperatures.[4] For lower temperatures, the charge transport is switched from nearest hopping to variable range hopping. By fitting the thermally activated transport equation, we extracted activation energy $E_a$ as a function of $V_{bg}$, as shown in Figure S4b. According to the point where $E_a$ deviates the linear trend, we estimate a Schottky barrier height of $\sim 35$ meV.

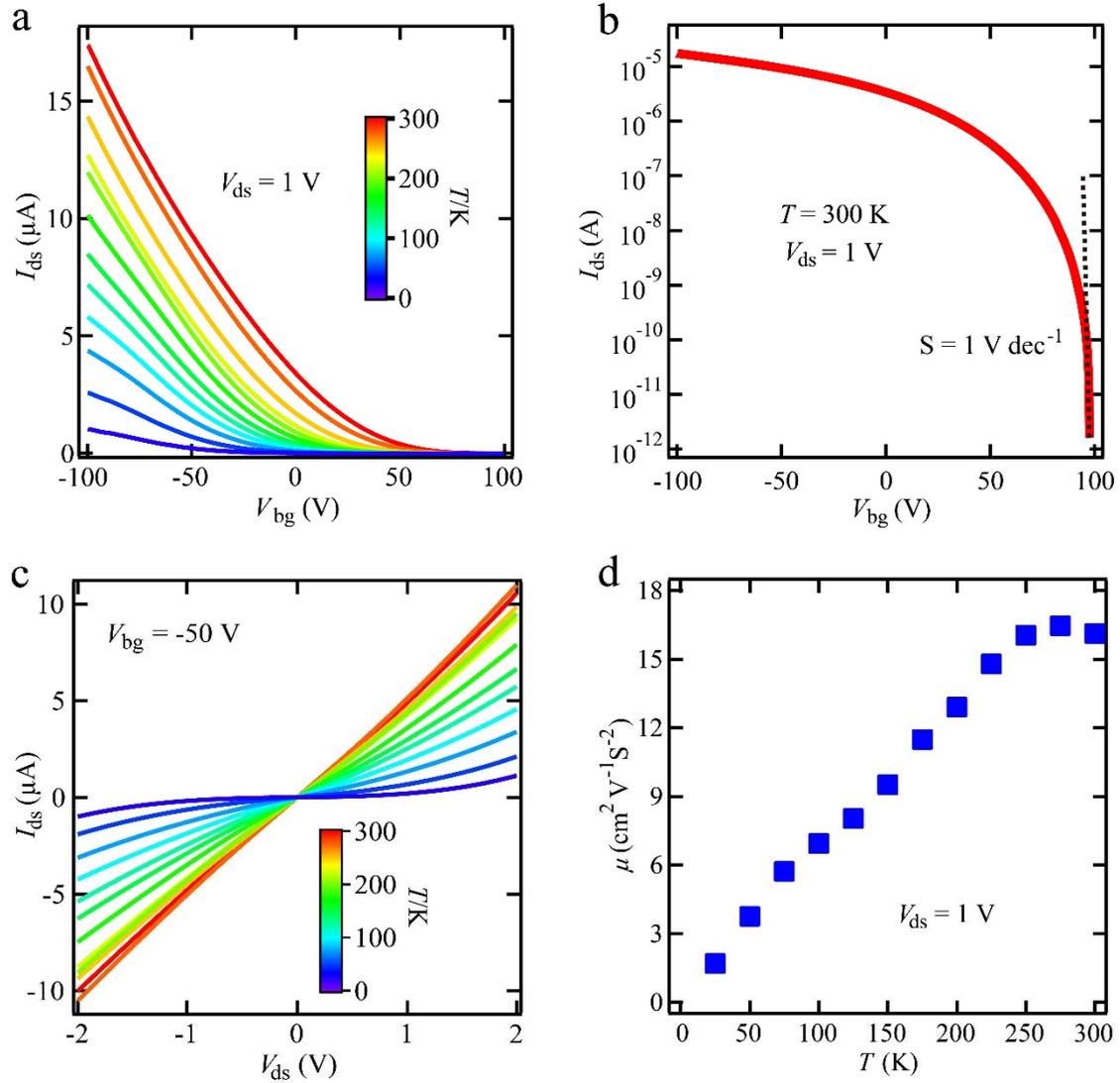

**Figure S3.** FETs properties of $Ta_{0.01}W_{0.99}Se_2$ few layer device. a) The transfer curves for $Ta_{0.01}W_{0.99}Se_2$ FET under various temperatures. The temperature changed from 25 K to 300 K with a step of 25 K. b) Logarithmic-scale transfer curve at room temperature (300K). Here, the extracted subthreshold swing is 1 V per decade. c) Temperature dependent $I_{ds}$-$V_{ds}$ curves at $V_{bg} = -50$ V, the nonlinear behavior vanished for temperatures above 100 K. d) The field-effect mobility ($\mu$) as a function of temperature, the highest mobility reached $\sim 16.5$ cm$^2$ V$^{-1}$ s$^{-1}$ at 275 K.

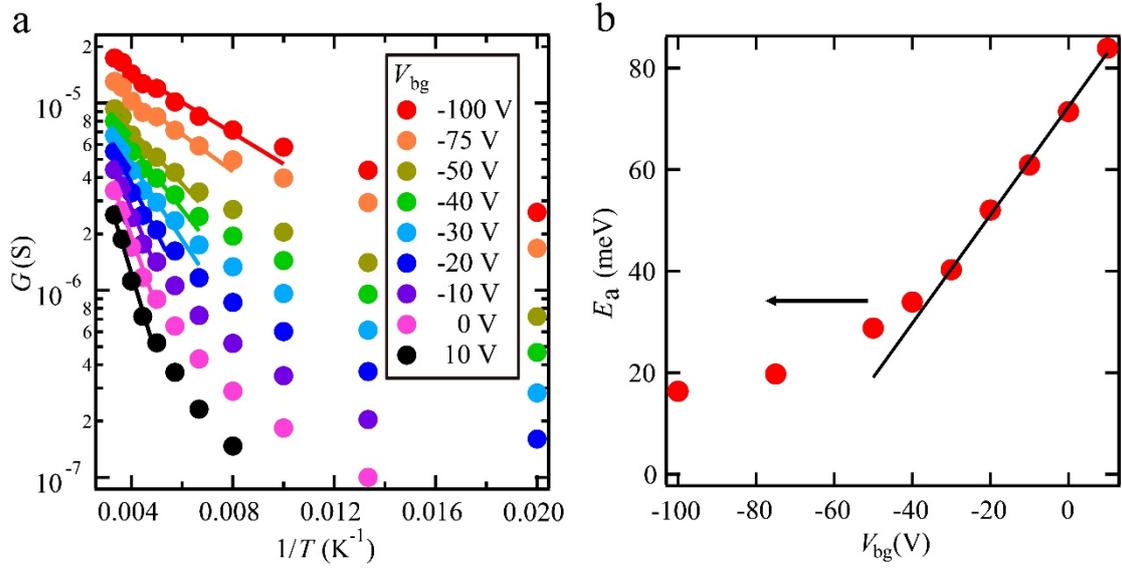

**Figure S4.** Thermally activated transport behavior of Ta-doped WSe$_2$ few layer device. a) Arrhenius plot of conductance, $G$, for different gate voltages. Solid lines are linear fits to the data showing activated behavior. b) Dependence of activation energies $E_a$ on $V_{bg}$. According to the deviation of the $E_a$ from the linear trend, we estimate a Schottky barrier height of ~ 35 meV.

**IV. FET property of pristine WSe₂ few layer device.**

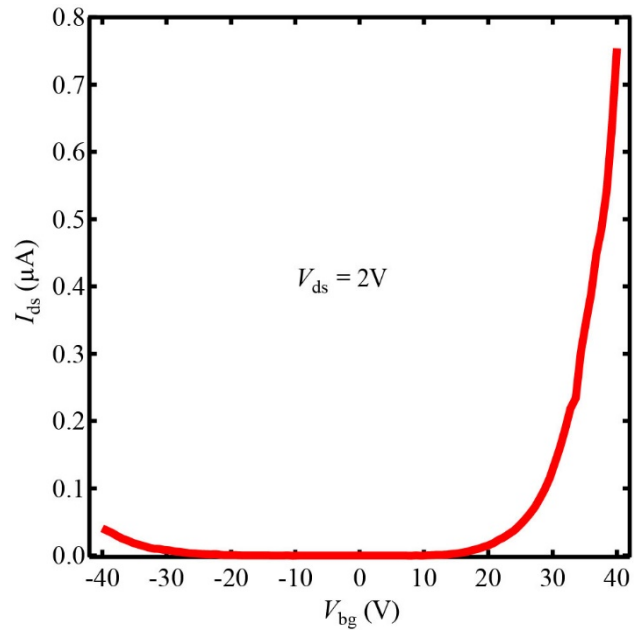

**Figure S5.** FET properties of a typical pristine WSe₂ few layer device (~3 nm). The contact electrodes were Ti/Au (5nm/40nm). An ambipolar behavior was observed, and the n-type branch was found to be more prominent.

## V. Photoresponsivity and external quantum efficiency of the $Ta_{0.01}W_{0.99}Se_2/MoS_2$ p-n heterojunction.

We calculated the photoresponsivity ($R$) and the external quantum efficiency (EQE) of the p-n heterojunction (the same device as shown in main text Figure 2) at $V_{ds} = 0$ V and $V_{bg} = 0$ V, as shown in **Figure S6**. Here, $R = I_P/P_{Laser}$ and EQE = $Rhc/(e\lambda)$, where $h$ is the Planck constant, $c$ is the speed of light, $e$ is the elementary charge, and $\lambda$ is the wavelength of the incident laser. Both $R$ (left axis) and the EQE (right axis) increase with the laser power decreases and reach values of up to ~95 mA $W^{-1}$ and ~19% ($P_{Laser} = 13.6$ μW), respectively. Which is comparable to the $WSe_2$ based vertical heterojunctions and outperforms the $WSe_2$ lateral p-n junctions.[5, 6]

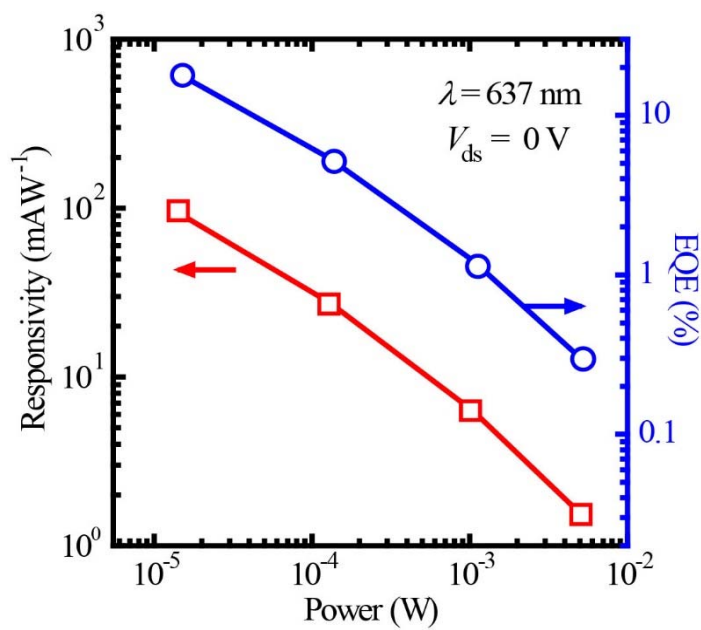

**Figure S6.** Photoresponsivity ($R$) and external quantum efficiency (EQE) versus illumination power intensity of the $Ta_{0.01}W_{0.99}Se_2/MoS_2$ p-n heterojunction shown in main text Figure 2, $V_{ds} = 0$V and $V_{bg} = 0$ V. Both $R$ and EQE decrease with power intensity increases.

## VI. Photoresponse of homojunction based on $Ta_{0.01}W_{0.99}Se_2$.

In addition to measure the photoresponse of $Ta_{0.01}W_{0.99}Se_2/MoS_2$ p-n heterojunctions, we further studied the performance of homojunction based on $Ta_{0.01}W_{0.99}Se_2$. Thanks to the continuous band alignments at the homojunction interface, the less carrier trap sites compared to heterojunction will contributed to photovoltaic response.[1, 7, 8] **Figure S7**a shows the optical micrograph of a typical $Ta_{0.01}W_{0.99}Se_2/WSe_2$ device. Pristine (2.9 nm) and Ta-doped WSe₂ ($Ta_{0.01}W_{0.99}Se_2$, 3.0 nm) are highlighted by green and white dashed lines, respectively. The fabricated homejunction exhibit impressive rectifying behavior at room temperature (inset of Figure S7b), since the junction formed at the interface. Then, we conducted the photoresponse measurement by exposing the device to a focused laser beam with various excitation powers (at a fixed laser wavelength $\lambda$ = 637 nm), as shown in Figure S7b. Nonzero $I_{ds}$ at zero bias under various laser power were detected, which indicate the existence of photovoltaic effect (similar to the heterojunction shown in main text). The calculated short circuit current ($I_{sc}$) and open circuit voltage ($V_{oc}$) as a function of incident laser power intensity ($P_{Laser}$) are shown in Figure S7c. Both $I_{sc}$ and $V_{oc}$ are found to increase with increasing $P_{Laser}$. Moreover, we calculated the photoresponsivity ($R$, left axis) and the external quantum efficiency (EQE, right axis) of the p-n homojunction at $V_{ds}$ = 0, 1, and 2 V (with $V_{bg}$ = 0 V), as shown in Figure S7d. Both $R$ and EQE are kept constant with different $P_{Laser}$ for $V_{ds}$ = 0 V. While for $V_{ds}$ = 1 V and 2 V, $R$ and EQE increase with $P_{Laser}$ decreases. The maximum value for $R$ and EQE are $\sim$ 482% and $\sim$ 2450 mA W$^{-1}$ respectively, at $V_{ds}$ = 2 V and $P_{Laser}$ = 1.36 µW. The outstanding photoresponse performance suggests that the Ta-doped WSe₂ holds high quality and has broad prospect in complementary electronic device applications.

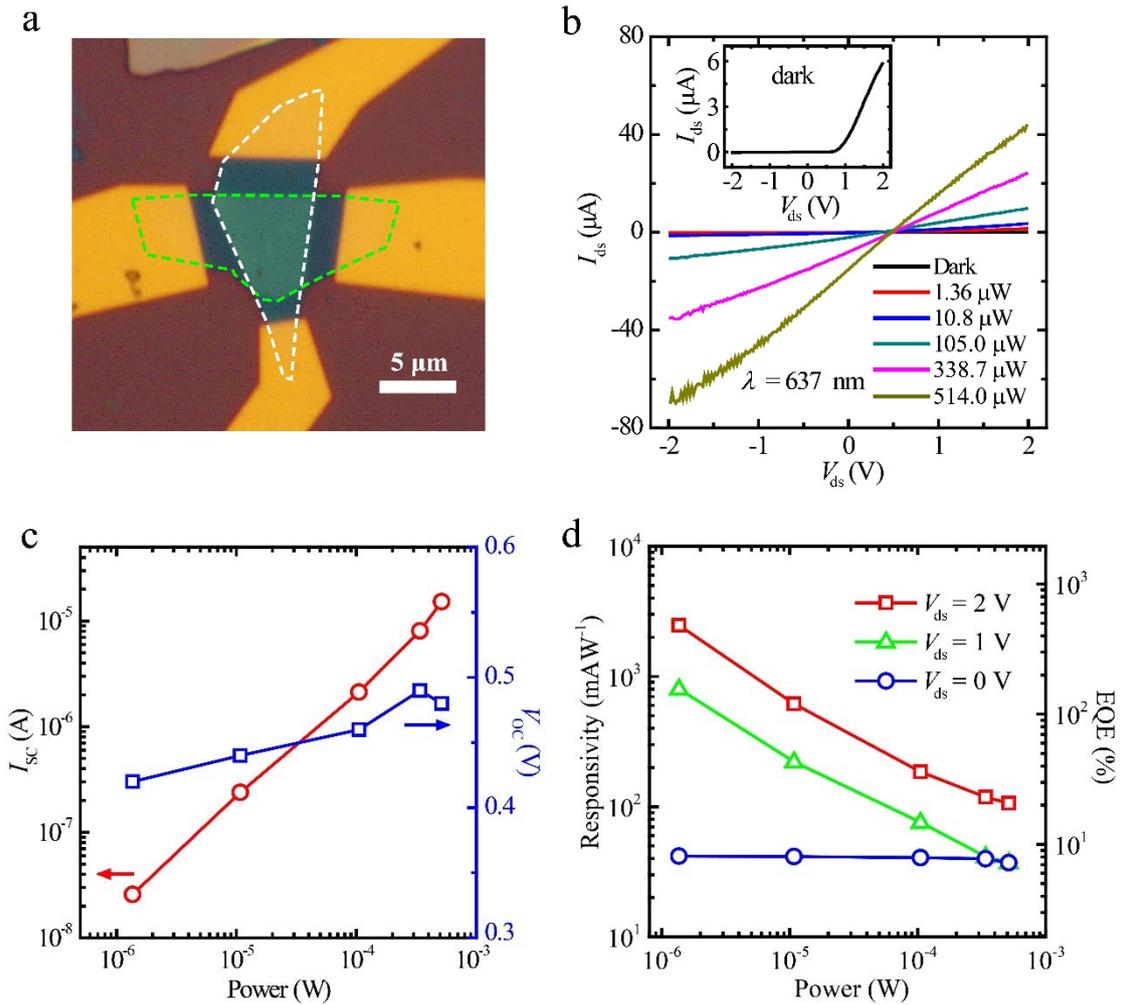

**Figure S7.** Photoresponse of a typical $Ta_{0.01}W_{0.99}Se_2$/WSe₂ homojunction. a) Optical micrograph of the measured device, the scale bar is 5 μm. Pristine and Ta-doped WSe₂ are highlighted by green and white dashed lines, respectively. b) $I_{ds}$-$V_{ds}$ curves measured by exposing the p-n diode to a focused laser beam (637 nm) with different power intensity. Inset: The rectifying behavior of the device at $P_{Laser} = 0$ and $V_{bg} = 0$ V. c) Short-circuit current ($I_{sc}$) and open-circuit voltage ($V_{oc}$) as a function of various incident power intensity. d) Photoresponsivity $R$ (left axis) and external quantum efficiency EQE (right axis) versus illumination power intensity at different $V_{ds}$, and $V_{bg} = 0$ V.